\documentclass[smallextended]{svjour3}       

\usepackage{amsfonts} 
\usepackage{graphicx} 

\usepackage{color}

\begin{document}

\title{Magnetic Resonance Imaging study of sheared granular matter}

\author{Jing Wang$^{1\ast}$ \and Zohreh Farmani$^{2\ast}$ \and Joshua Dijksman$^2$ \and Cindy L\"ubeck$^3$ \and Oliver Speck$^{1,3}$ \and Ralf Stannarius$^1$}
\institute{
Jing Wang,
\email{jing.wang@ovgu.de} \\
 Zohreh Farmani,
\email{zohreh.farmani@wur.nl} \\
 Joshua Dijksman,
\email{joshua dijksman@wur.nl} \\
 Cindy L\"ubeck,
 \email{cindy.luebeck@ovgu.de} \\
 Oliver Speck,
 \email{oliver.speck@ovgu.de} \\
Ralf Stannarius
\email{ralf.stannarius@ovgu.de}\\  \smallskip
$^\ast$ shared first authorship
\\
$^1$  Institute of Physics, Otto von Guericke University Magdeburg, D39106 Magdeburg, Universit\"atsplatz 2, Germany\\
$^2$~Physical Chemistry and Soft Matter, Wageningen University \& Research, Stippeneng 4, 6708 WE Wageningen, The Netherlands\\
$^3$~Forschungscampus Stimulate, D-39106 Magdeburg, Otto-Hahn-Stra\ss{}e 2, Germany }

\titlerunning{MRI of sheared granular matter}       
\authorrunning{Wang et al.}     

\date{}

\maketitle
\begin{abstract}
We introduce a Magnetic Resonance Imaging technique to study the geometry of shear zones of soft, low-frictional and hard, frictional granular materials and their mixtures. Hydrogel spheres serve as the soft, low-frictional material component, while mustard seeds represent rigid, frictional grains. Some of the hydrogel spheres are doped with CuSO$_4$ salt to serve as tracers. A
split-bottom shear cell is sheared stepwise and the shear profiles are determined from the differences of tomograms after
successive shear steps, using Particle Imaging Velocimetry. We find that the shear zone geo\-metry  differs considerably between soft grains
submersed in water and the same material without the embedding fluid.
\end{abstract}

 \maketitle

 \section{Introduction}

Granular matter responds to shear stress with a behavior quite different from that of liquids. In contrast to simple liquids
which distribute the shear uniformly, granular matter develops shear bands, where the shear deformation is usually localized
in narrow zones. The geometry of such zones depends on the material properties and the shear geometry in a complex way.
A characterization of the grain displacements inside a sheared granular bed is thus of high interest, from a practical point
of view as well as from the aspect of fundamental understanding of granulate physics. The shear characteristics of granular
matter is also relevant in natural phenomena, e.g. in the formation of avalanches or the stability of geological faults.
One of the interesting phenomena
is the tendency to reduce the packing fraction in shear zones, described by Reynolds \cite{Reynolds1886} already in 1886,
and first studied quantitatively by Bagnold \cite{Bagnold1954,Bagnold1966}.

In the present study, we will focus on the
question whether ensembles of soft, slippery grains behave quantitatively or even qualitatively different from
hard, frictional grains when they are sheared in the same geometries.
The detection of shear zones inside 3D granular beds is often a nontrivial task. With few exceptions, granular matter
is opaque and optical methods fail to visualize the reorganization of the grains in the bulk. When one wants to
gain access to the internal processes in a granular bed and at the same time avoid destructive methods such as
excavation, a couple of imaging techniques are available \cite{Amon2017}. In transparent material, one can use optical fluorescence
imaging with a laser sheet  \cite{Tsai2003,Tsai2004,Panaitescu2012,Dijksman2017}.

While this technique is rather affordable and simpler than the competing tomographic methods, it has several limitations.
One restriction is that the material has to be submersed in an index-matching fluid, and this may influence the shear
characteristics.
A much more elaborate and expensive imaging technique is X-ray Computed Tomography. One can scan a variety of materials
with this technique, and scanners are available with different spatial resolutions from micrometer sizes to centimeters
\cite{Richard2003,Zhang2006,Borzsonyi2012,Wegner2012,Wegner2014,Szabo2014,Stannarius2019,Stannarius2019b}. A recent review by Weis and Schr\"oter can be found in Ref.~\cite{Weis2017}.
This technique is quite efficient when one can resolve the individual particles and perform Particle Tracking (PTV) or Particle
Imaging (PIV) Velocimetry. However, it is relatively complicated to label individual particles as tracers, even though this is in principle
possible. The problem is that X-ray absorption and particle densities are rather closely related to each other. If one chooses
tracers with the same size as the background material, but with different X-ray absorption, one usually has to deal with tracers of
different weights and sedimentation may become an issue. A workaround can be hollow particles as tracers in a bed of filled
particles with the same weight and outer diameter.

Magnetic Resonance Imaging (MRI) is an alternative method that exploits the nuclear magnetization of the material for imaging  \cite{Sederman2007,Stannarius2017}. MRI-visible objects can be imaged directly, if they contain liquids e.~g. \cite{Nakagawa1993,Nakagawa1994,Sommier2001,Penn2019,Penn2020},
or NMR-invisible grains can be immersed in an MRI-visible fluid, e.~g. \cite{Finger2006,Naji2009,Fischer2009}, so that they appear as 'shadows' in the tomograms.
In the present study, we expand the opportunities of MR Imaging by mixing Magnetic Resonance visible grains with magnetically invisible grains of practically identical mechanical properties. This is achieved by a special preparation technique described in Sec.~\ref{sec:material}.

While MRI has a number of drawbacks compared to X-ray CT, one of the advantages is the opportunity to have tracers with practically the
same physical properties as the bulk material except for the MRI visibility. This can be achieved either by using liquid-filled vs.
solid particles, or, as in our experiments, by doping an amount of tracers. This technique allows us to monitor the displacement of the material
inside the granular bed with PIV or PTV.
Nuclear magnetic resonance can be used to reveal the inner structure of a granular bed either by packing density profiles \cite{Nakagawa1993,Sakaie2008}, or even at the grain level \cite{Fischer2009,Borzsonyi2011}.
Both MRI and X-ray CT are comparably slow methods, so that typically real-time imaging is not possible. There are some
exceptionally fast techniques that allow the real-time imaging at least of slices inside the granular bed, for example with MRI
\cite{Penn2019,Penn2020,Engel2020} and X-ray CT \cite{Bieberle2016,Waktola2018,Stannarius2019b}.

In the present study, we analyse displacements of grains in shear experiments. With the equipment and imaging method used in this study, the time to record one tomogram is about one minute. During the data acquisition, the granular ensemble has to be at rest. Thus, the sample is sheared in steps and MRI tomograms are recorded after each step in the static sample.

Two principal geometries can be employed to
study shear of granular materials.
Linear shear cells have a simple geometry, but they limit the displacement and are therefore suited only for oscillatory shear \cite{Panaitescu2010b,Panaitescu2012}. In one study, ling linear sliders were used to determine the distortion and reflexion of shear zones at the boundaries between layers of different granular materials \cite{Borzsonyi2011}.
For continuous shear experiments, closed shear zones in rotating systems are more suitable. Plate-plate or cone-plate geometries are preferentially used,
where rheological problems like the shear rate dependence of the material is a focus \cite{Kuwano2013,Fall2015}.
For the investigation of both transient and asymptotic states of shear zones,
cylindrical split-bottom containers have proven advantageous
\cite{Fenistein2003,Fenistein2004,Sakaie2008,Cheng2006,Fenistein2006}.
An inner, rotating granular volume is sheared relative to an outer, fixed volume.
Earlier investigations have dealt with the geometry of the shear zone  \cite{Fenistein2003,Fenistein2004,Unger2004,Unger2007,Fenistein2006,Cheng2006,Dijksman2010}, dilatation of the shear
zone \cite{Sakaie2008,Wegner2014}, alignment of shape-anisotropic grains under shear \cite{Wegner2012,Wegner2014},
or the formation of surface heaps and secondary flows \cite{Wortel2015,Fischer2016}.

These previous shear investigations exclusively dealt with frictional, hard particles. In the present study, we investigate the role of
softness and slipperiness on the shear characteristics.
We employ hydrogel spheres which are characterized by a very weak elastic
modulus of the order of 50 \dots 100 kPa and a friction coefficient that is more than one order of magnitude smaller than for typical
granular materials\cite{Ashour2017b}.
It has been demonstrated earlier that during silo discharge, hydrogel spheres behave qualitatively different from hard, frictional grains. Some features resemble rather liquid behavior that that of hard grains, for example, the pressure is almost hydrostatic over a substantial granular bed height. In addition to the pure material, we prepare mixtures of these soft grains with hard, frictional spheres in order
to quantify the influence of the mixing ratio on the shear characteristics. We test whether the shear characteristics changes
qualitatively when the admixture of hard, frictional spheres reaches a certain percolation threshold so that chains of these particles traverse
the sheared regions.

\section{Experimental setup and materials}
\subsection{Setup geometry}

\begin{figure}[htbp]
\centering
 \includegraphics[width=0.66\columnwidth]{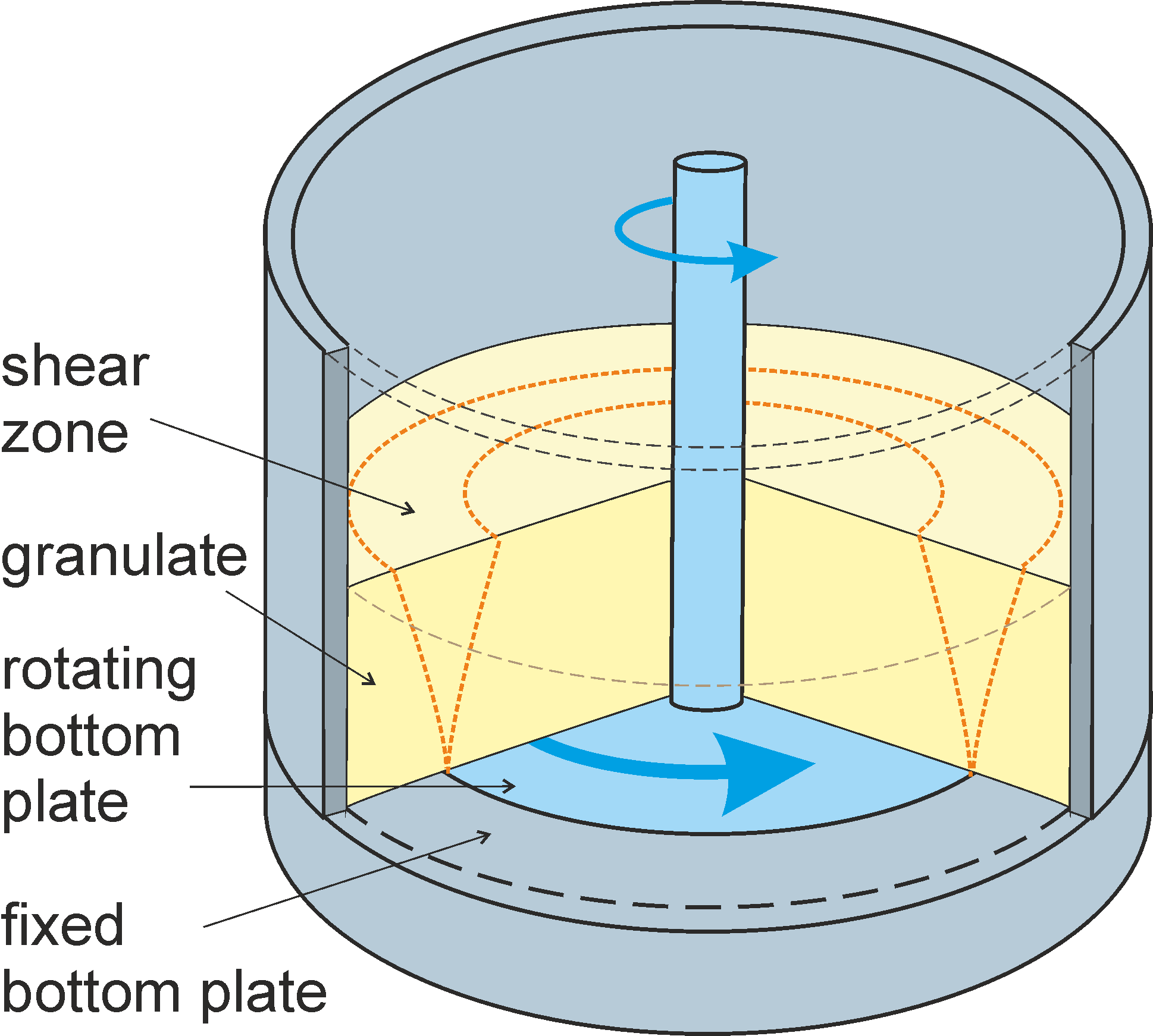}
  \caption{Sketch of the experimental setup: Grey parts are fixed in the MRI scanner, while
  bottom disk and central rod sketched in light-blue can be rotated about a vertical axis by well-defined
  steps. The granular material in the outer parts remains static and fixed to the outer container,
  while in the axial part it rotates with the bottom disk.
  The bottom surface was roughened in order to prevent the first layer to slide relative to the borders.
  The dashed lines sketch the boundaries of the shear zone.}
  \label{fig:setup}
\end{figure}

In order to keep the setup small enough to be inserted in the MRI scanner, the decision was made to use a split-bottom cylindrical shear geometry.
 Figure \ref{fig:setup}
is a sketch of the cell used in our experiment. It is manufactured from acryl glass (Perspex) which is MRI invisible. The inner disk can be rotated by hand from the top, via the rod that connects to the bottom disk. Its rotation angle is measured by a goniometer. In addition, two
MRI-visible tracers are glued to the bottom of this disk so that the position and rotation state of the
container can be directly retrieved from the MRI tomograms. The inner diameter of the container
is $D_0=2 R_0=150$~mm, and the rotating disk has a diameter of $D = 2R_s=95$~mm.

The geometry and width of the shear zone depends on the ratio $H/R_s$ of the filling height $H$ and the radius $R_s$ of the
rotating bottom plate
\cite{Unger2004,Unger2007,Cheng2006,Fenistein2004,Fenistein2006,Sakaie2008}. At small fill heights, ($H<0.6~R_s$), it has been found for rigid, frictional spherical grains that the shear zone passes
the granular bed completely as shown in Fig.~\ref{fig:setup} and the inner part of the granular material rotates synchronously
with the bottom. When the fill height becomes larger and the axial rod shown in Fig.~\ref{fig:setup} is missing, the shear zone closes in and at sufficient heights it forms a closed dome
above the bottom disk. In our experiment, the geometry is slightly different (albeit, as we will show, with little consequences on the shear profile) because of the narrow
central rod that is used to turn the bottom disk.

The MRI scanner used was a 3 Tesla Siemens MAGNETOM Skyra at the Stimulate Forschungscampus Magdeburg with 70~cm bore size and short magnet. The tomograms have a voxel resolution of $0.75 \times 0.75 \times 0.7$~mm$^3$. Recording of a single
tomogram takes about 1 minute, and the setup is kept in a static condition during the recording. The acquisition time $\tau$ was  of the order of a few ms, which is very short compared to the transversal relaxation time $T_2>1$ s of pure water. $T_1$ of water is about 3~s in
our experiments.

\subsection{Material}
\label{sec:material}

The soft, low-frictional material used in our experiments are hydrogel spheres (HGS). Two sizes of hydrogel spheres were studied, small ones and large ones (see Tab.~\ref{tbl:material}).
The dry powder of the small HGS (type ``snow'') was received from {\em JRM Chemical} in a polydisperse mixture with diameters from a few micrometers to about 0.5 mm. The dry grains were
sieved to obtain a narrow size distribution between 0.3~mm and 0.355~mm. After swelling in distilled water, the grains reached diameters between 2~mm and 3~mm.

In some experiments, we employed large HGS. The dry particles were obtained from
a commercial supplier ({\em Happy Store, Nanjing}). They were swollen in a mixture of distilled water and NaCl.
The salt concentration determines the final diameter of the hydrogels. It was chosen such that the diameter was approximately 7~mm,
with a polydispersity of about 5~\%.

In addition, we used mustard seeds as hard, frictional grains in mixtures with the small HGS of comparable size. In principle, the swollen hydrogel spheres as well as
the mustard seed contain liquids and provide a $^1$H NMR signal. However, because of the fast repetition rate of scans in our experiments, the
recovery of the nuclear magnetization between subsequent scans is very small and the signal is weak due to the long $T_1$. This is a desired effect in our experiment, since we are interested to (i) keep the experiment time short and (ii) use
specially prepared tracers. The tracers are hydrogel particles that are swelled with an 8~mM aqueous solution of CuSO$_4$. The copper sulfate reduces the longitudinal relaxation time $T_1$ of the nuclear magnetization substantially \cite{Kjaer1987}.
Thus, the hydrogel particles swollen in the dopant solution produce a strong MRI signal. On average, we have used a share of
about 20~\% of doped hydrogels spheres in the experiments if not stated otherwise.

\begin{table}[htbp]
\small
  \caption{\ List of materials used}
  \label{tbl:material}
  \begin{tabular}{|lll|}
    \hline
    Material & mean diameter  & mean weight  \\
    & [mm]& [mg]\\
    \hline
    large hydrogel spheres & 7 & 180 \\
    small hydrogel spheres & 2.5 & 8  \\
    mustard seeds & 3 & 15 \\
    \hline
  \end{tabular}
\end{table}

\subsection{Data evaluation}

\begin{figure}[htbp]
\center
 \includegraphics[width=1.0\columnwidth]{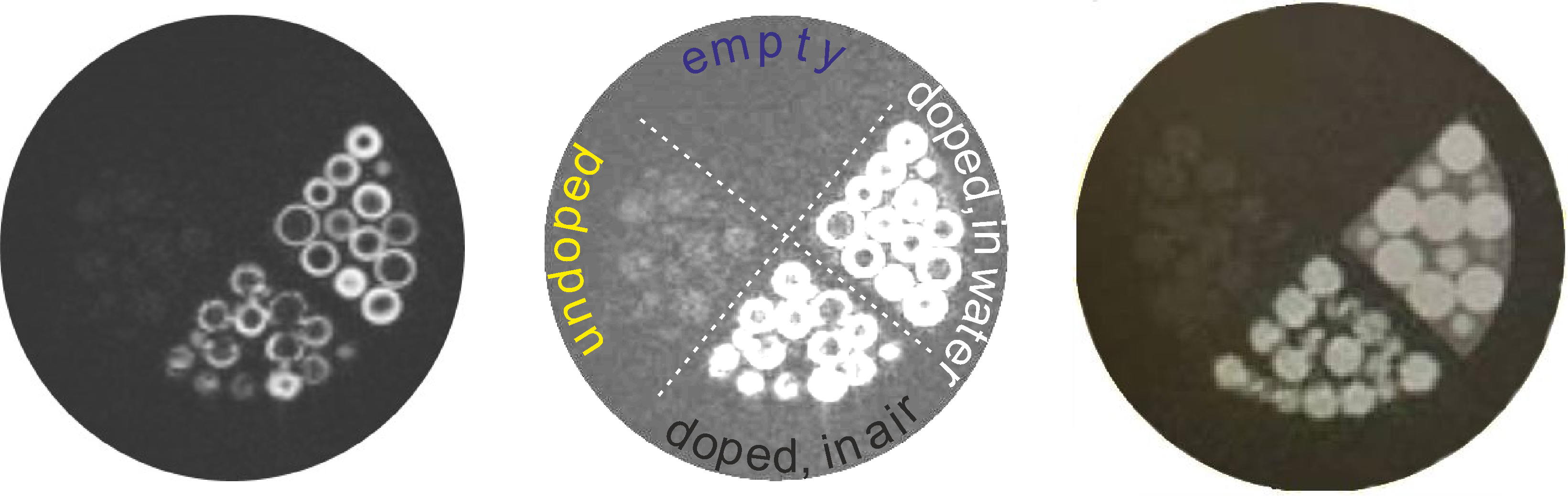}
   \caption{Left: Horizontal slice of the tomogram of a four compartment cylindrical plastic container (50 mm inner diameter) with large (7~mm) hydrogel spheres.
  The top sector is empty. The left sector contains non-doped spheres, swollen in NaCl solution.
  The bottom sector contains spheres doped with copper sulfate, and the right sector is filled with
  the same spheres submerged in distilled water. Middle: strongly contrast enhanced image of the same
  slice, which also shows the signal from undoped HGS (in the left sector).
  Right: Same sample as in the top right image,
  after 5 months. The CuSO$_4$ doped spheres are now completely visible, they appear as solid spheres in the tomograms, not as shells}
  \label{fig:largeHGS}
\end{figure}

We observed that the swelling of the large HGS in CuSO$_4$ solution led to an unexpected effect right after swelling: the outer shells (about 1-2 mm)
of the 7 mm hydrogel spheres became visible in the MRI tomograms, while the inner part was swollen but remained
invisible in the tomograms (Fig.~\ref{fig:largeHGS}).
These particles can be easily tracked in subsequent tomograms of sheared samples. For the small HGS,
the chosen spatial resolution of the tomograms is at the limits when individual spheres are traced,  but with a PIV technique it is easy to track the displacement of local particle configurations.

Figure \ref{fig:largeHGS} shows a slice of a test tomogram recorded with large hydrogel grains. The view is a horizontal cross section of
a cylindrical container with four compartments. The sector on the left is filled with 7 mm HGS swollen in
NaCl solution. The NMR signal is very weak because $T_1$ is much larger than the repetition rate of the
scans, so that the nuclear magnetization does not recover between the scans. In the contrast enhanced image (middle), their weak signal is seen. In the bottom sector of the Figure, 7~mm HGS doped with copper sulfate
are contained. They provide a strong NMR signal in their outer shell. The selected slice cuts the hydrogels spheres in different positions,
therefore the rings seen in the slice have different diameters. This shell-type tomographic pattern of single spheres remained even
after one week. The copper sulfate requires several weeks to diffuse into the sphere center and to provide an MRI signal of filled spheres with our technique. Note also that, since the HGS in the bottom compartment were surrounded by air, the susceptibility mismatch led to artifacts that let the spheres appear rugged.
The individual spheres are nevertheless clearly separated. In the right sector, the same doped spheres were submerged in distilled
water. This removes the susceptibility mismatch artifacts completely. Each grain can be clearly identified.

Figure \ref{fig:largeHGS} (right) shows the same sample after several months. The spheres appear filled now. Obviously, the copper sulfate has diffused inside the spheres. Within the present study, we have not tested the dynamics of this process. It requires between a few weeks and several months.

\begin{figure}[htbp]
\centering
 \includegraphics[width=0.9\columnwidth]{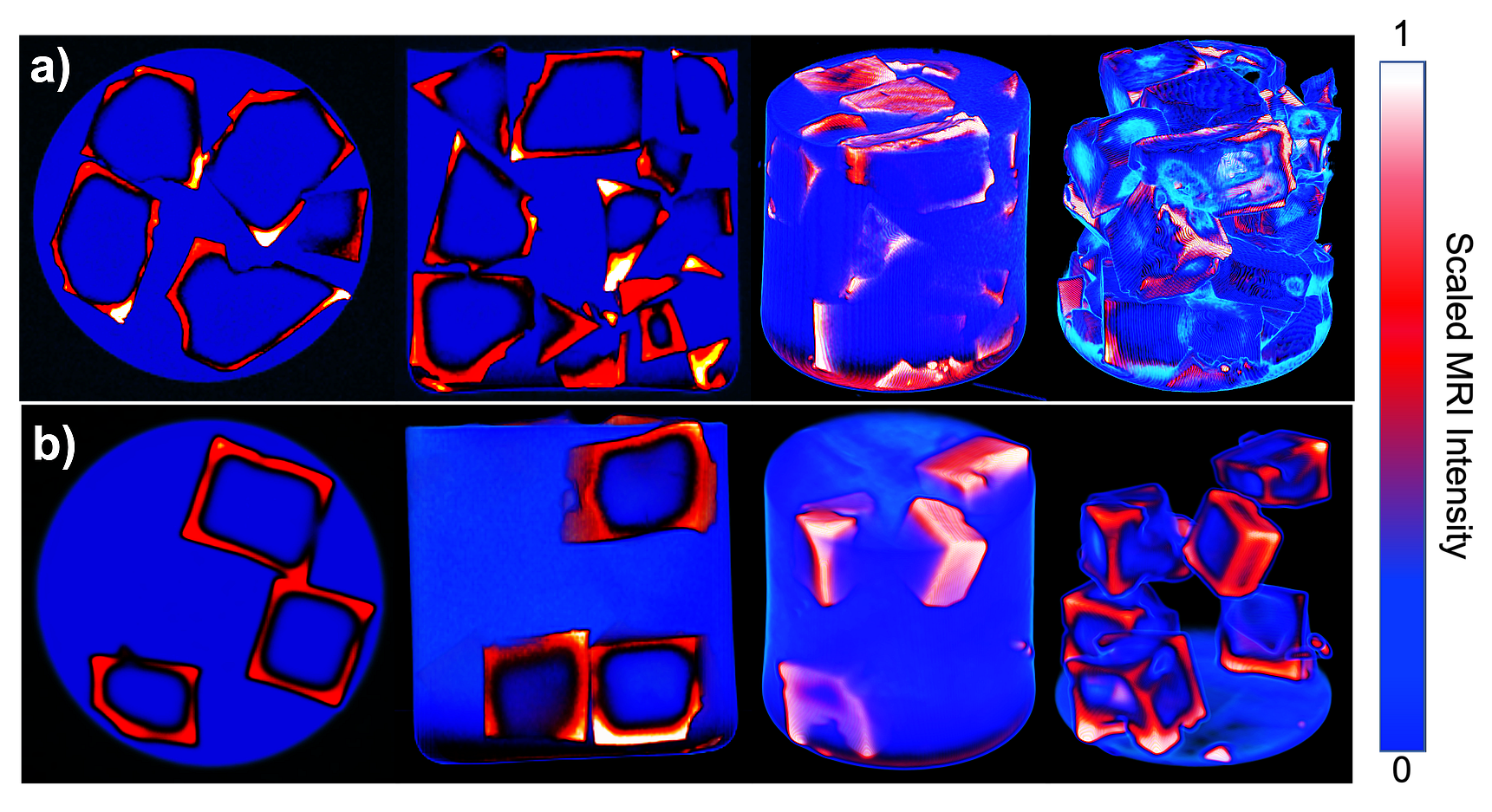}
  \caption{Nearly cube-shaped hydrogel blocks freshly swelled in 8~mM copper sulfate solution. (a) all hydrogel blocks labeled with CuSO$_4$, (b) only 30\% of the blocks labeled.
  From left to right we  show a central 2D horizontal slice, a 2D central vertical slice, 3D rendering and 3D rendering with water outside the cubes removed.
  Red to white regions label thin surface regions of the particles. Blue regions (low NMR signal) are further inside the blocks and around them. The container is a cylinder of 100 mm diameter, filled to a height of $\approx 95$ mm. MRI intensities are scaled by the maximum value.}
  \label{fig:cubeHGS}
\end{figure}

Before we describe our shear experiments with hydrogel spheres in detail, we note that the slow diffusion of the CuSO$_4$ into hydrogel particle cores during the swelling process provides a unique tool when one is interested to scan ensembles of irregular grains. During the first days after the swelling process, only the outer borders of the grains are NMR-visible. In order to demonstrate this, We scanned centimeter-sized irregular, but nearly cube-shaped hydrogel blocks. Different percentages of them (100\% and 30\%) were 8~mM CuSO$_4$ doped. The swelling process took 24 hours, and the MRI scans were performed 5 days after swelling. The 2D cross sections of tomograms in Fig.~\ref{fig:cubeHGS} confirm that the CuSO$_4$ decorates only zones at the surface of the blocks, in a similar way as the spherical hydrogels (Fig.~\ref{fig:largeHGS}). This makes the individual blocks visible in the scans. If the complete hydrogel material would be NMR visible, blocks lying side-by-side would hardly be distingushable. When we decrease the share of doped particles in the packing (Fig.~\ref{fig:cubeHGS}b) we only see the doped blocks (30\%). One can easily track the doped blocks by segmenting the doped ones, as they are visible through edges. We conclude that this method can be extended
successfully to any shape of hydrogel particles.

\begin{figure}[htbp]
\centering
 \includegraphics[width=0.75\columnwidth]{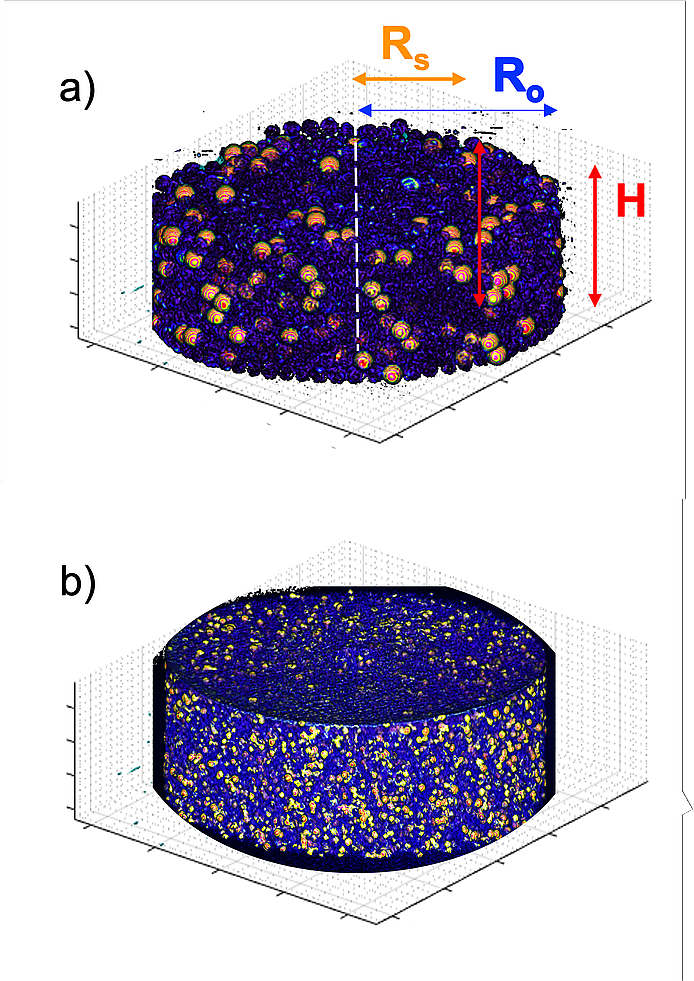}
   \caption{The figure shows typical examples of reconstructed 3D images of large (a) and small (b) hydrogel sphere ensembles. Copper sulfate doped particles appear bright in the dark (blue) background. The sample size is 15~cm in diameter.}
  \label{fig:zohreh1}
\end{figure}

Typical reconstructions of the grains in our shear cell are shown in Fig.~\ref{fig:zohreh1}. Image (a) is a reconstruction of the large HGS ensemble with the doped tracers bright(yellow) and the non-doped background particles dark (blue). Image (b) shows the small
HGS with the bright tracers and a dark background of voids and non-doped spheres.

In the shear experiments with large hydrogel spheres, we employ PTV to determine shear zone profiles from the traces of the doped particles.The concentration of these tracers is low enough so that they can be easily identified and collated in subsequent tomograms.
The bottom plate was rotated in steps of $\approx 30$ degrees, and the tracer displacements between these shear steps were measured. To monitor the actual rotation angle, two tracers were glued below the rotating bottom plate. Their positions in the tomograms were used to confirm the correct position of the setup in the scanner and to determine the exact actual rotation angle of the bottom disk.

The assumptions implicitly made in this approach are that (1) the shear zone has axial symmetry around the central axis, and (2) that the initial transients after filling the container have ceased. The measurements show that after approximately 30 degrees of rotation of the bottom disk, a stationary profile of the shear zone is established. The 3D tomograms were divided into rings with constant height $h$ above the bottom and constant distance $r$ from the central axis. We average over $5\times 5$ tomogram pixels (approximately $3.5\times 3.5$~mm$^2$ in the $(h,r)$ plane.
In each ring, the tangential displacement of doped grains was determined. The angular displacements of the tracers were scaled with the rotation angle of the bottom disk.
Each experiment consisted of 7 to 11 evaluated shear steps of about 30 degrees. The data of these steps were averaged for each given fill height. In order to avoid influences of a transient behavior, the initial shear step was excluded.
The advantage of this method is that the displacements can be determined with high precision for each doped grain, a disadvantage is the comparably poor spatial resolution of this method which is of the order of the HGS particle diameters.
The choice of 30 degree rotation steps is a compromise. Too small rotation steps reduce the accuracy of the displacement vectors, while too large rotation steps can lead to a loss of particles between subsequent tomograms in a given ring, when the particle changes its
radial or vertical position.

With the small grains, the same experimental steps
were performed: stepwise slow rotations of the bottom plate by 10$^\circ$. In these experiments, we did not follow individual grains but employed PIV to retrieve displacements in different zones of the granular bed. This results in a better spatial resolution
because of the smaller particle diameters. The tomogram data are sampled in a ring of constant height above the bottom plate and distance from the rotation axis, and the correlations of the profiles along this ring are evaluated.

\section{Results}

\subsection {PTV of large grains}
\begin{figure}[htbp]
\centering
\includegraphics[width=0.85\columnwidth]{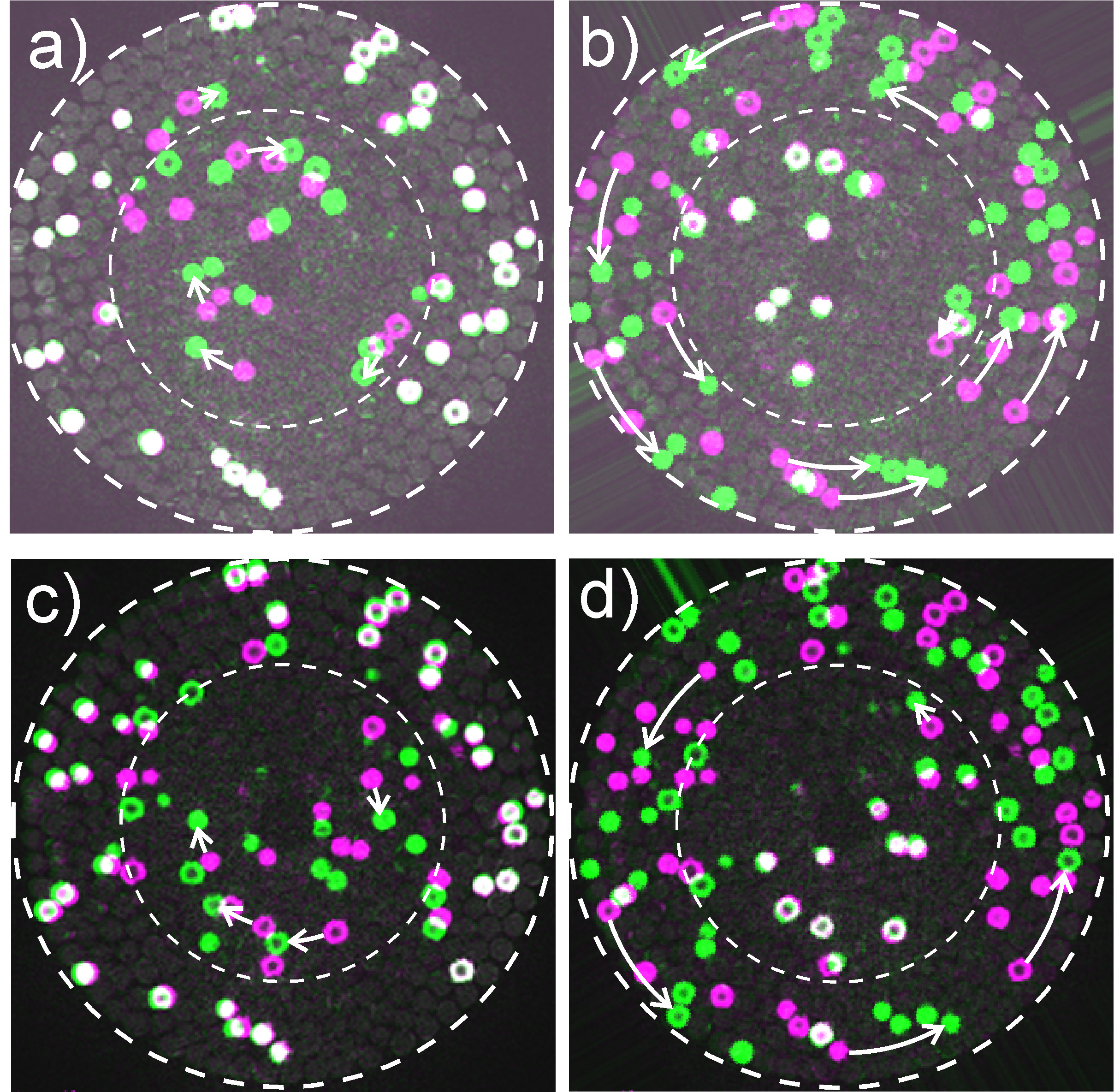}
  \caption{Displacement of large HGS during $\approx 30$ deg rotations of the bottom plate. A horizontal slice at  10~mm height within a 20 mm high granular layer was selected. The dashed lines mark the
  inner of the container wall and the border of the rotating bottom disk. (a) shows in white the immobile tracers, in magenta the tracer positions after filling, and in green the displaced positions after rotation of the plate. Arrows indicate examples of selected displacements of individual tracers.
   (b) shows the differences in the frame of the rotating disk
    (disk rotation compensated by appropriate digital rotation of the slice), white tracers moved with the disk,
   magenta and green are tracers in the 0$^\circ$ and 27$^\circ$ tomograms, resp.
   (c,d) present the same operations with the same slice between the 210$^\circ$ and 240$^\circ$ rotations of
   the bottom plate.}
  \label{fig:schnitte}
\end{figure}

Figure  \ref{fig:schnitte} was constructed from horizontal slices of tomograms recorded in a 20 mm high layer of
large HGS in our setup. One tomogram was taken after filling the container, then, the inner disk was rotated by nominally 30$^\circ$ and a second tomogram was recorded. From the displacement of the markers at the bottom of the rotating disk, the actual rotation angle was established to be precisely only 27$^\circ$. The rotation speed was chosen sufficiently slow so that the sample was sheared quasi-statically. We can tacitly assume that the ratio of the angular displacement of a tracer at given radius $r$ and height $h$ and the angular displacement of the bottom disk is equal to the ratio $\omega/\omega_0$ of the angular velocities during slow, continuous shearing.
The first two images (a,b) show the overlay of identical slices
of the two tomograms in pseudocolor: The green color plane of one slice was replaced by the green color plane of the second slice, while the red and blue color planes remained unchanged. White spots in image (a) indicate tracer spheres that have not changed their positions, dark grey spots represent the background of non-doped spheres and light grey (green and magenta) spots are particles that have been displaced. They mark the differences of both tomograms. Positions of tracers in the 0$^\circ$ tomograms are reproduced in magenta, those in the 27$^\circ$ tomogram in green. For selected grains, the mapping of grain positions in the two tomograms are indicated by arrows that represent the displacement. In (a), the difference between the original, non-rotated tomograms is shown. The outer grains  remained at their original positions, while the grains in the center have rotated with the bottom disk. The inner dashed circle marks the position of the edge of the rotating disk. The slice shown is located at $h=10$~mm above the bottom, in the middle of the granular bed.

In (b), we have digitally rotated the second tomogram backwards by the rotation angle of the bottom disk. Now, white stands for particles that co-rotate with the bottom disk. Indeed, the marked particles in the inner part coincide, which indicates that they rotated in the laboratory frame by the same angle as the bottom disk. The bottom two slices (c,d) in the figure
shows the same overlays with tomograms recorded at $210^\circ$ and $240^\circ$ rotation angles. At these rotation angles, we expect that a stationary shear zone has established.

\begin{figure*}[htbp]
\includegraphics[width=\textwidth]{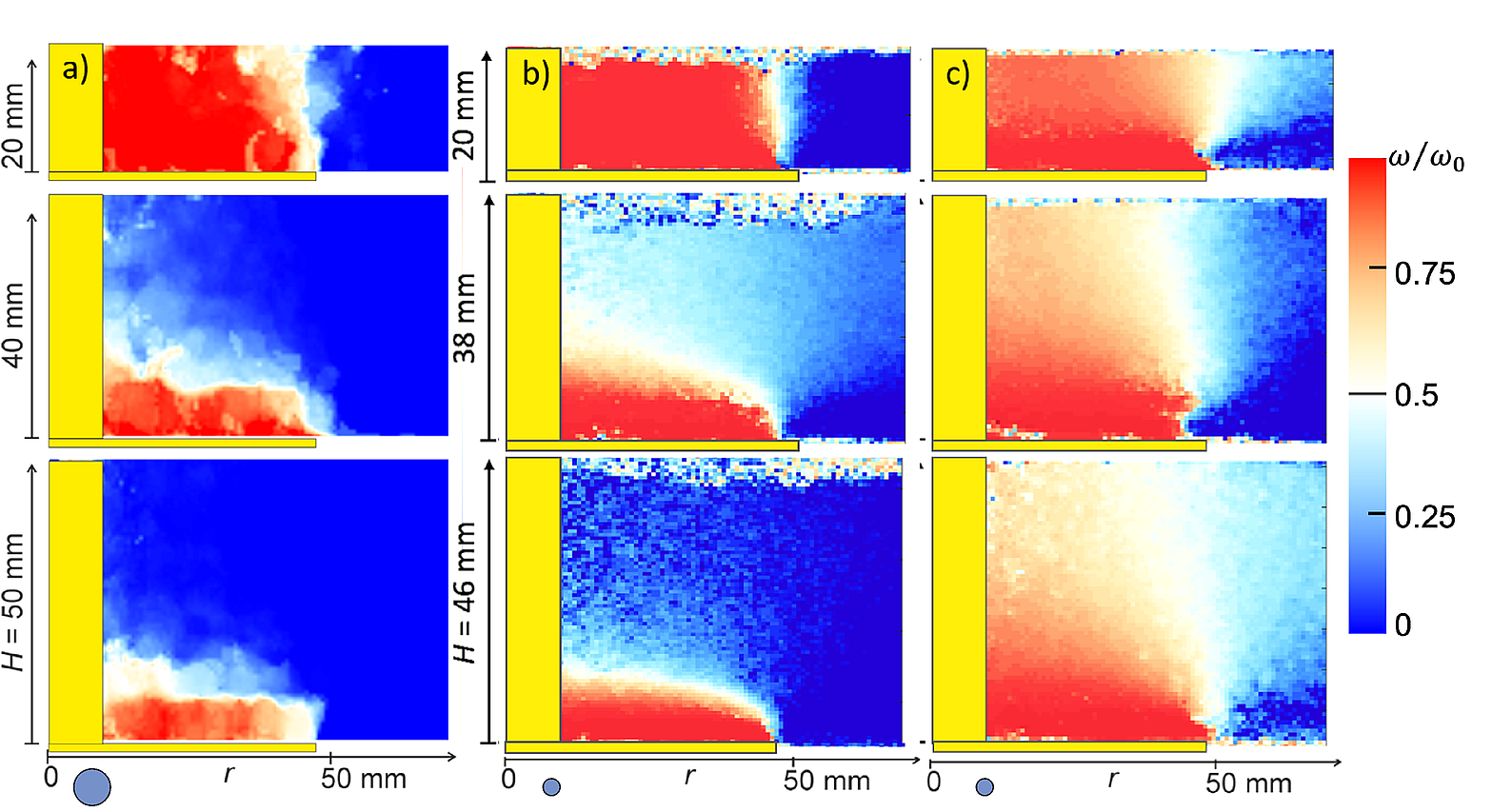}
\caption{Flow profiles, a) for large HGS (without immersion fluid), b) for small HGS (without immersion fluid), and c) for small HGS (grains immersed in water) in beds of different heights. Grey circles sketch the typical bead sizes. The vertical and horizontal bars indicate the positions of the central rod and the bottom disk, respectively. The angular velocity $\omega(r,h)$ was normalized by the bottom disk angular velocity $\omega_0$. The shear zone where the angular velocity ranges between $0$ and $\omega_0$ is visible as a bright band separating the grains at rest (outside) and co-rotating with the disk (center).}
\label{fig:shear3}
\end{figure*}

The complete shear profile can be obtained by following the tracer positions in each ring of a given distance $r$
from the axis and height $h$ above the bottom by means of PTV, on a particle level. The result is shown in Fig.~\ref{fig:shear3}a.
Here, we have determined the displacements of all marked grains
in rings around the central axis, with a resolution of approximately one particle radius.
The displacement angles were scaled with the rotation angle of the
bottom disk. If one assumes that the ratio of angular displacements in a given height and radius is proportional to the rotation angle of the bottom disk, one can extrapolate the results to the effect of a continuous rotation.
The ratio of rotation angles to the angle of the bottom disk can be assumed to be equal to the ratio of rotation speeds in a given $(h,r)$ zone to the rotation frequency $\omega_0$ of the bottom disk. In Fig.~\ref{fig:shear3}a,
results for three different fill heights of the container were
compiled, The top image is for a bed height
$H\approx 20$~mm ($H/R_s = 0.42$), the middle part for
$H\approx 40$~mm ($H/R_s = 0.84$), and the bottom image for
$H \approx 50$~mm ($H/R_s = 1.05$).
The first layer of grains above the bottom disk rotates
with that disk in all cases. In the shallow granular bed ($H=20$~mm) the shear gradient is almost completely horizontal everywhere, with a slight widening of the shear zone towards the top. With increasing bed height, the shear zone bends inwards (middle image of Fig.~\ref{fig:shear3}a). For the largest bed height studied in our MRI experiments ($H=50$~mm), the shear zone is nearly vertical everywhere, confined to the first layer of HGS. Above a height of about 4 HGS diameters, the shear zone is confined to a narrow shell around the central rod that serves as the rotation axis of the bottom disk.
In general, the PTV method is rather inaccurate for small $r$, near the central axis, because the displacement of tracers is very small there.
Nevertheless, it is obvious from the tomogram data that
the particles near the pole essentially follow the motion of the outer layers, the pole has little influence. This means, in particular, that for sufficiently large fill heights the closed dome-shape of the shear zone is almost the same as the one expected without a pole, with only small deviations in the lower part of the pole, near the horizontal dome.
It is also noticeable that the shear zone for large fill heights above the bottom disk becomes rather sharp. This is a consequence of some lattice-like induced positional order of the spheres above the disk, in flat layers. Obviously, there is a sharp drop of the co-rotation with the bottom disk already above the second layer.

\begin{figure}[htbp]
\centering
a)\includegraphics[width=0.7\columnwidth]{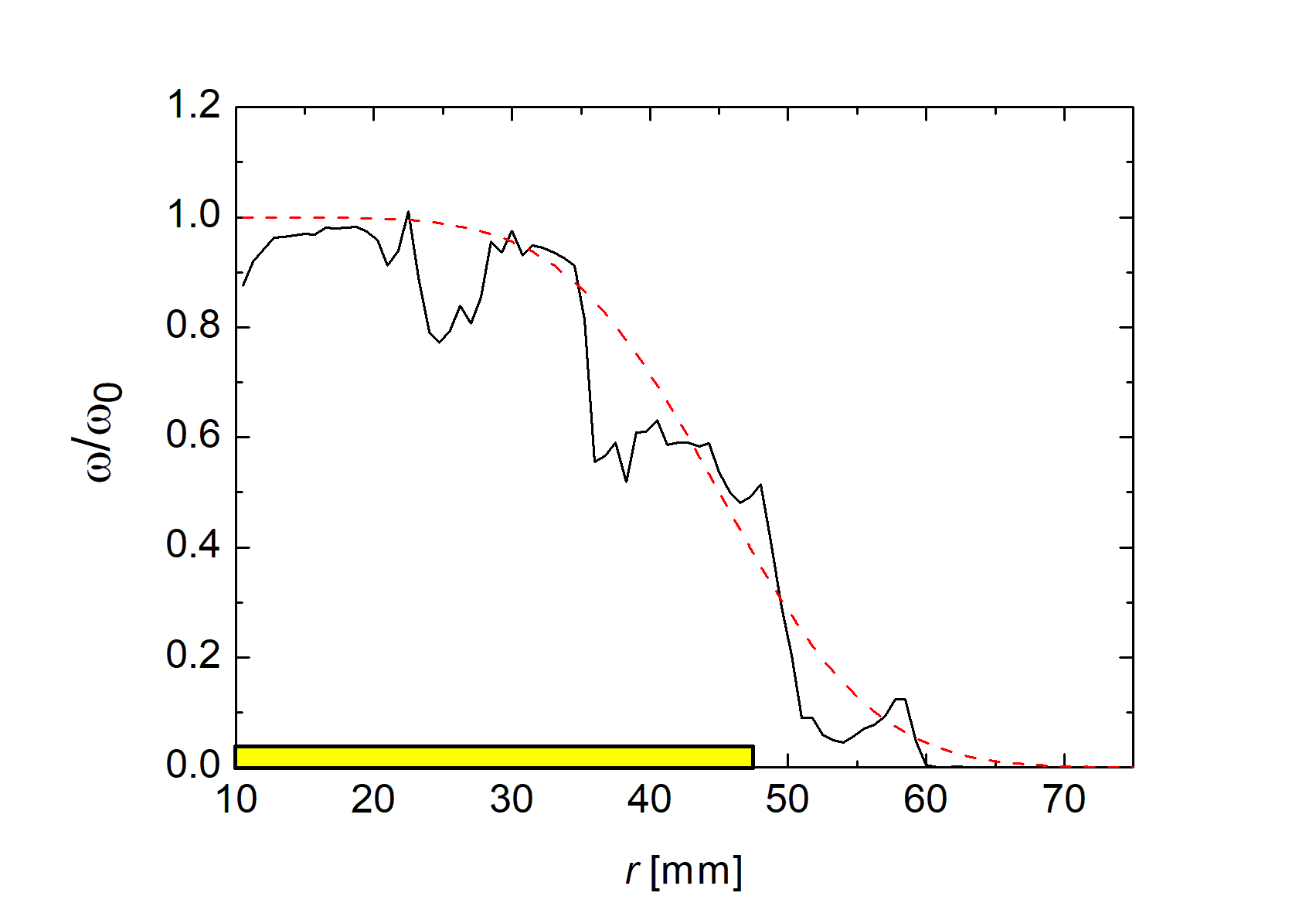}\\
b)\includegraphics[width=0.7\columnwidth]{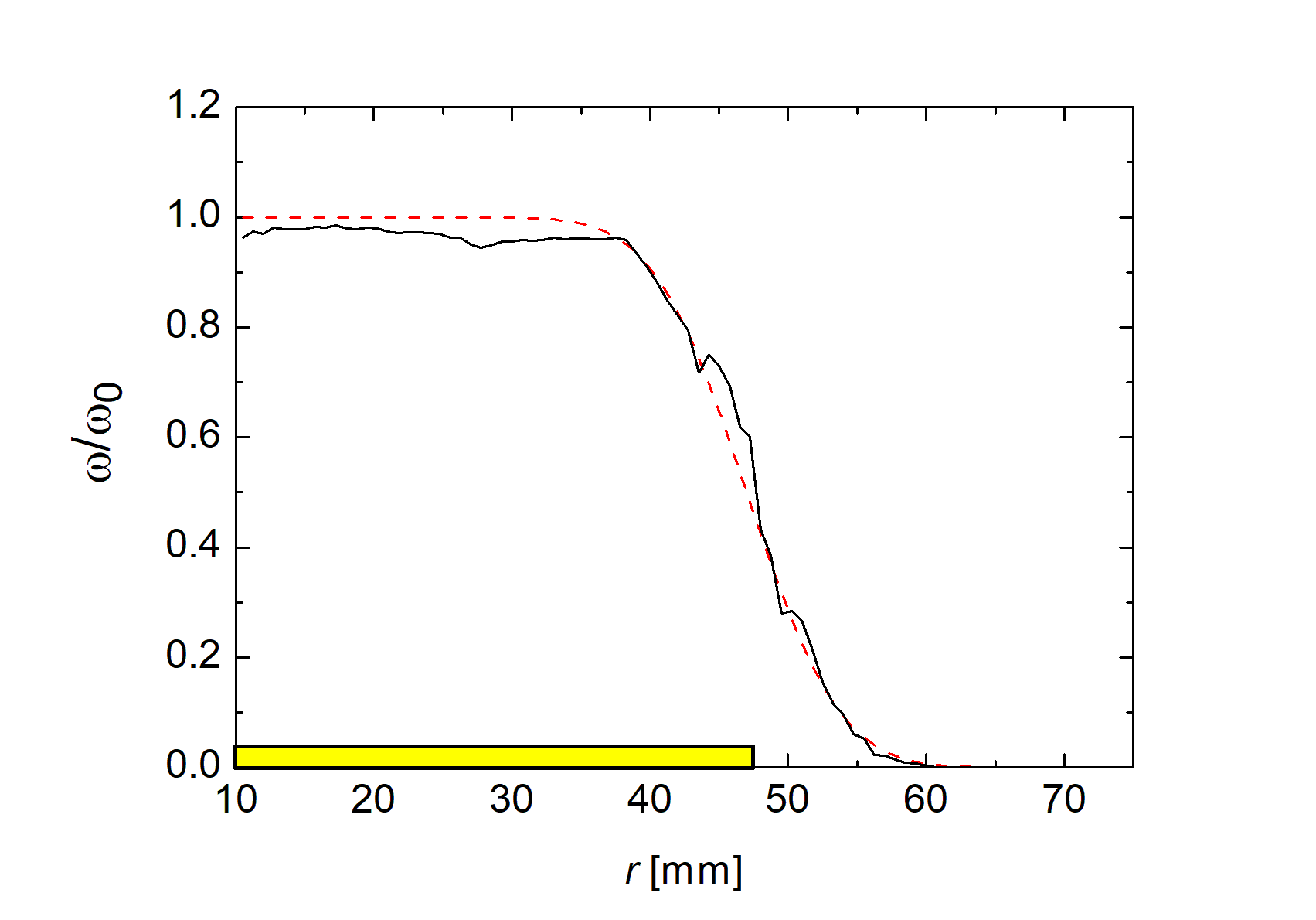}
  \caption{Shear profiles extracted from the $H=20$~mm experiment
  with large grains. The graphs show the profiles (relative rotation speed in a ring of radius $r$ at height $h$ for
$h=H$ (a) and $h=H/2$ (b). The profiles are averages of 6 angular difference measurements each. The parameters for the erfc-graphs (dashed lines) are given in the text.}
  \label{fig:largeprofiles}
\end{figure}
It is possible to establish the quantitative shape of the shear zone from these data, and Fig. \ref{fig:largeprofiles} shows two examples. In the upper graph, the horizontal cut through the shear zone in the top layer is seen, a slice of five voxels height (3.75~mm) was averaged. The profile was fitted with a scaled complementary error function
\begin{equation}
\label{ERFC}
\omega(r) = \frac{\omega_0}{2}\mbox{erfc}(r) = \frac{\omega_0}{\sqrt{\pi}}\int_r^\infty \exp{(-x^2)} dx,
\end{equation}
using $x=(r-R_c)/\delta$ with the shear zone center radius $R_c$ and width $\delta$. The parameters are $R_C=45$~mm, $\delta =12.5$~mm in the top layer and $R_c=47$~mm, $\delta= 7.5$~mm
in the central layer. Note that $R_c$ is quite close to the bottom disk radius, as expected for this low fill height. This empirical function captures the profile satisfactorily well, and the parameters $\delta$ and $R_c$ may be used for a straightforward characterization of the shear zone width and position, in order to compare of the influences of material parameters on the shear zone geometry. In more accurate experiments, we have focused on smaller spheres that are not equally well distinguished as single particles in our tomograms, and we have used PIV for the reconstruction of the shear profiles.

\subsection{PIV of small grains}

The study of the small HGS was performed in two parts. In the first part, the HGS were filled into the container after the
water at their surfaces had been removed as thoroughly as possible. Still, some excess water attached to the spheres was present, but in very small quantity, so that the voids between the spheres were filled almost exclusively with air. Then, the same experiments as with the large grains were repeated. We will call this condition dry environment. The results are shown in Sec.~\ref{sec:dry}. In another set of experiments, we filled the voids between the HGS in the shear cell with water. We will refer to this situation as wet environment. Results will be discussed in Sec. \ref{sec:wet}.

\subsubsection{Hydrogel spheres in air}
\label{sec:dry}

Figure Fig.~\ref{fig:shear3}b shows the results for small HGS in the same fashion as Fig.~\ref{fig:shear3}a for the large HGS.
It is obvious that the results look very similar to the large particle PTV results, with a slightly better spatial resolution in the PIV results for small HGS.

The PIV produces some artifacts in the surface layers of the granular bed. The reason is that small height differences in different regions are incorrectly interpreted by the correlation technique.
If one disregards the upper 5 mm, there is a clear picture of the shear zone geometry: At low fill height, the shear zone reaches the top, and it is almost vertical, with a slight inclination inward towards the top layer. Above a certain fill height, the shear zone closes. The height of the dome-shaped shear zone approaches a saturation value. When $H\approx R_s$, it reaches about $0.2~R_s$. These results are qualitatively similar to results of earlier studies of rigid, frictional grains.

\subsubsection{Hydrogel spheres immersed in liquid}
\label{sec:wet}

In the second set of experiments, tomograms were recorded with submersed HGS. The density of the HGS is a few percent higher than that of water so that the buoyancy compensates approximately 95~\% of the weight. This reduction of the effective weight also compensates the contact forces between the grains. Yet this is not the only effect that has to be considered: The water filling the voids between the small grains removes capillary forces that presumably exist in the dry environment because the HGS surfaces are not completely dry, and liquid bridges can form between touching spheres. Thus, one may expect a reduced friction force. Finally, the embedding water has to be redistributed when the grain ensemble is sheared. This involves the inner friction of the fluid and may lead to a more liquid-like shear characteristics in our material.

The MRI results are shown in Fig.~\ref{fig:shear3}c. The presence of interstitial water does not affect the MRI spectra because it gives an equally weak signal as the HGS swollen in pure water. The shear zone is much broader for all fill heights. As in the dry experiment, the central region that rotates with the bottom disk extends to the top layer at heights up to $H\approx 0.4R_s$, but then, with increasing bed height, the shear zone does not close dome-like above the disk. Even when the bed height $H$ is comparable to the disk radius, the top layers react to the disk rotation, although with a reduced rate.
The shear profile is qualitatively different from that of the dry material. This is in a way somewhat counter-intuitive, since a reduced effective weight leads to lower vertical contact forces and one would probably assume that the transmission of shear forces in the vertical direction is reduced.

\subsubsection{Mixtures of hard frictional and soft, low-frictional grains}

\begin{figure}[htbp]
\centering
\includegraphics[width=0.9\columnwidth]{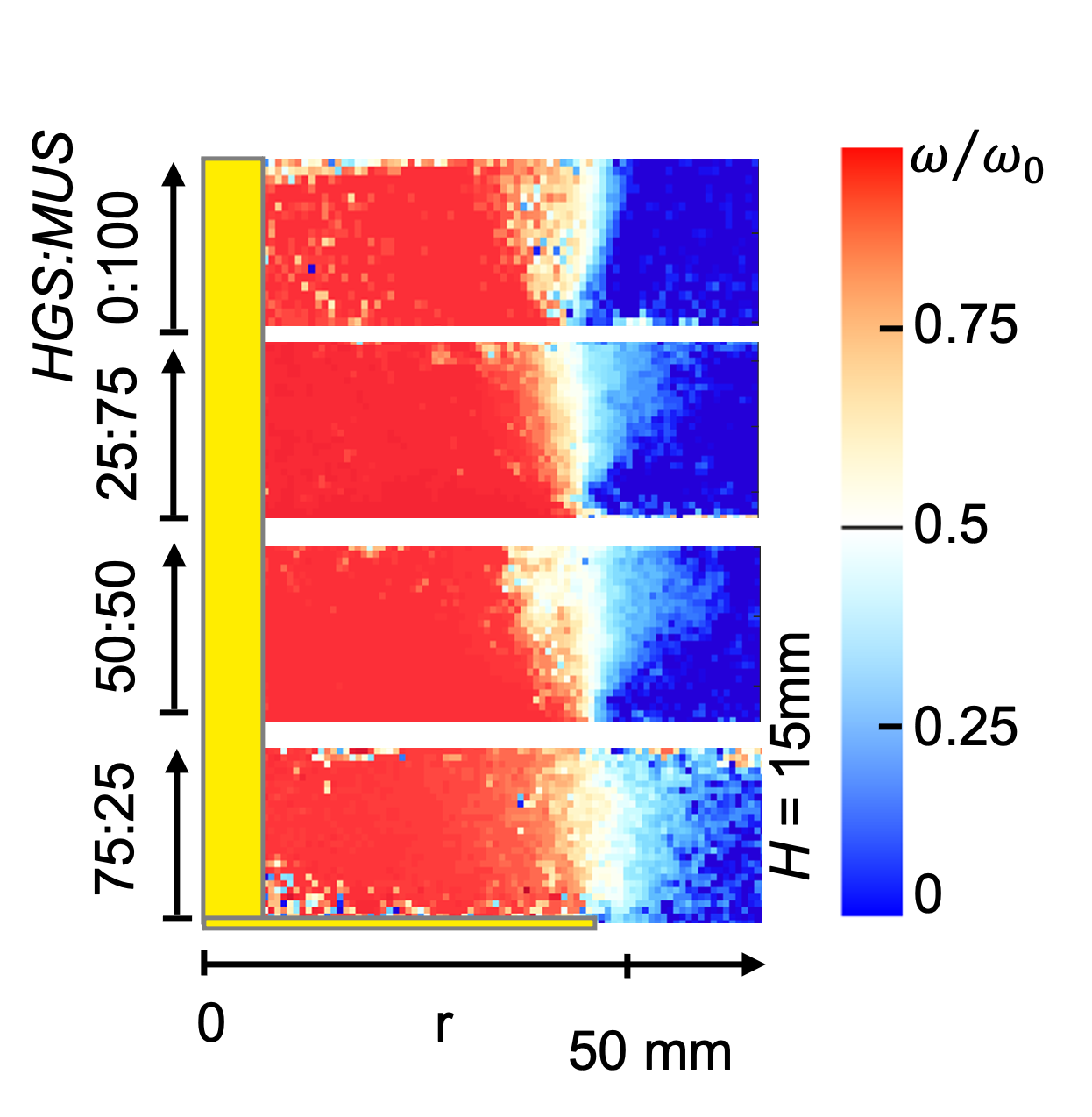}
  \caption{Shear zone profiles for small HGS mixed with rigid, frictional mustard grains in different concentrations. The vertical and horizontal bars indicate the positions of the central rod and of the bottom disk, respectively. The plot shows the ratio of the displacement angles of grains at ($h,r$) and scaled by the rotation of the bottom disk. No immersion fluid was present.}
  \label{fig:mix}
\end{figure}

In the third set of experiments,
the HGS were mixed with rigid frictional mustard seeds of a comparable sizes in different mixing ratios.
Here, the experiments were restricted to a single fill height, 15~mm. Within our experimental limits, we cannot see a clear difference
in the flow profiles of the rigid pure mustard
seed sample and the pure soft, low-frictional HGS sample (see Figs. \ref{fig:shear3}a,b and \ref{fig:mix}, top). This is, at first glance, an astonishing observation in view of the fact that the outflow of HGS
from hoppers has been found to be qualitatively different from that of hard, frictional grains \cite{Harth2020,Wang2021a}. We will discuss the differences between the two experiments below in Sec. \ref{sec:concl2}.
On the other hand, the shear zones of the mixtures are wider than those of the pure materials. This could be an indication that introducing "soft spots" in a bed of hard grains may increase fluctuations and the propagation of strain

\section{Conclusions}

\subsection{Methodical achievements}

We have demonstrated that shear profiles in soft particle ensembles can be measured by means of a non-invasive Magnetic Resonance Imaging approach. The displacement of particles was determined from differences of tomograms after stepwise rotation of the bottom disk in a split-bottom container. Two techniques have been employed to extract the particle displacements during the discrete shear steps.
The first method was used for the large HGS. The positions of individual marked tracers in given heights $h$ and radii $r$ were evaluated and a displacement profile was derived. The method requires that in each ring ($h,r$), at least one tracer is found. This condition was fulfilled with tracer concentrations between 10 \% and 20 \%. The method is comparable to an earlier investigation of shear profiles in linear sliders \cite{Borzsonyi2011} where poppy seeds distributed in beds of glass spheres and corundum grains were tracked. The advantages of the method applied in this study over the earlier system are twofold. First, we can strictly exclude segregation in the pure HGS systems, since the tracers are identical in their physical properties to the non-marked background particles. Second, doping with copper sulfate substantially reduces the recording time of the spectra.

The second approach, PIV, was used for the small HGS. The correlations of tomogram profiles in rings of constant $(h,r)$
provided the mean displacements in each ring, which were used to construct the displacement profiles. This method averages over the whole ring and it is therefore less suited to probe possible variations of the shear zone along an azimuthal cut. However, it provides a much better spatial resolution than the PTV of large grains.

Of course, PTV can in principle also be used in small particle ensembles (e.g. Ref.~\cite{Borzsonyi2011}) as long as the individual grains can be identified. Then, however, much smaller shear steps (of the order of 5$^\circ$ rotations) are needed for a correct mapping of the individual grains. Then, one either loses relative accuracy of the angular
displacements or much more tomograms need to be recorded.

We note that the investigation of hydrogel spheres of different sizes offers an, although limited, access to study the role of softness: The choice of particle diameters does not change the elastic modulus significantly, but the particle weight and thus gravitational as well as shear forces increase with approximately the third power of the radius. Thus the hydrogel spheres behave more like hard, low-friction grains in small size experiments but more like soft, deformable low-frictional grains in larger size experiments, when the setup is appropriately scaled with the particle sizes.

An additional interesting feature of the specific labeling method
of hydrogel particles that has been used here is the initial concentration of the CuSO$_4$ in the outer layers of the particles, irrespective of their particular shapes. The thickness of these layers is of the order of 1~mm within the first few days after swelling. This allows us to map particles by their outer borders. This feature is very helpful particularly when neighboring particles (e. g. polyhedra) touch each other with flat faces. If the labeling would make the complete particle visible in the tomogram, then such particles would be very hard to separate in the image data.

\subsection{Shear zone characterization}
\label{sec:concl2}

An important result of this study is the characterization of the shear profiles in ensembles of soft, slippery materials. We find that there is little difference between the profiles in mustard seed ensembles, as representatives of hard, frictional spheres, and HGS ensembles (in dry environment) representing  soft grains with very low friction. Mixtures of soft and rigid spheres do not show a clear trend. 

At first glance, this finding of similar shear zone profiles of the HGS and mustard seeds contradicts findings in hopper outflow, where the flow in a quasi two-dimensional container was found to
differ considerably between soft slippery, rigid frictional, and mixed systems. It was argued earlier that neither the softness nor the low friction alone cause the specific flow behavior \cite{Harth2020,Wang2021,Wang2021a,Pongo2021}, but a combination of both.

When comparing these results to the present experiment, one has to keep in mind that the gravitational pressure in the shear setup is considerably smaller than in the hoppers
studied the above cited studies. The granular bed is much shallower in this MRI study.
Thus, the softness of the spheres plays a
less important role in the shear experiment. In the hopper, the bottom pressure is of the order of several kPa, and it can deform the spheres by up to 10\% of their dimensions. In contrast, the gravitational pressure in the present shear cell hardly reaches 0.5 kPa, which has much lower effect on shape deformations of the HGS with elastic moduli of the order of 50 kPa. The other characteristic property of the HGS, their extremely low friction coefficient ($\mu$ of the order of 0.03)
is, however, equally effective in the shear experiment reported here as in the earlier hopper experiments. Thus, the comparison of the flow characteristics of small HGS in a shallow bed with flow of larger HGS in higher beds can help to discriminate the influences of friction and elasticity.

At the contacts between HGS, liquid capillary bridges may form, and can in principle influence the shear behavior. However, a good argument to neglect these bridges is the comparison of Figs.~\ref{fig:shear3}a and b. There is little difference between the shear profiles
of both kinds of particles. The masses of the particles in the study, however, relate like 22:1. If
capillary forces would influence the shear profiles, they would affect the small HGS much more than the large ones, and differences would become evident.

When the HGS in dry environment are compared with the submerged particles in the same geometries (Figs.~\ref{fig:shear3}b and c), considerable differences are evident. Since we have excluded the effect of capillary forces, the differences must be attributed to the shear of the interstitial liquid. The system of low-frictional HGS with interstitial fluid behaves more liquid-like, i.e. the shear zone tends to spreads across the cell. This behavior deserves a more detailed investigation in future experiments. It should be noted that we do not expect a similar effect with immersed frictional grains, at least at sufficiently low shear rates, because there, the shear forces will be primarily mediated by particle-particle contacts. With the low-frictional HGS, however, internal friction within the liquid may be of comparable influence as the direct particle contacts
A dependence of the ratio of these frictional forces on shear rates is expected. This should be done in rheological experiments by shear-rate dependent torque measurements.

Finally, the comparison of hard, frictional mustard seeds and soft HGS shows that both systems behave rather similar in dry environment. The mixtures of hard and soft grains appear to show a tendency to broaden the shear zones, which needs further exploration.

\section*{Conflicts of interest}
There are no conflicts to declare.

\section*{Acknowledgements}
David Fischer, Torsten Trittel and Christoph Klopp are cordially acknowledged for important contributions to the construction of the setup.
This project has received funding from the European Union's Horizon 2020 research and innovation programme under the Marie Sk\l{}odowska-Curie
grant agreement {\sc CALIPER} No 812638.


\begin{thebibliography}{10}

\bibitem{Reynolds1886}
O.~Reynolds.
\newblock Experiments showing dilatancy, a property of granular material,
  possibly connected with gravitation.
\newblock {\em Proc. Roy. Inst. Great Britain}, 2:354--363, 1886.

\bibitem{Bagnold1954}
R.~A. Bagnold.
\newblock Experiments on a gravity-free dispersion of large solid spheres in a
  newtonian fluid under shear.
\newblock {\em Proc. Roy. Soc. A}, 225:49, 1954.

\bibitem{Bagnold1966}
R.~A. Bagnold.
\newblock The shear and dilatation of dry sand and the 'singing' mechanism.
\newblock {\em Proc. Roy. Soc. A}, 295:219, 1966.

\bibitem{Amon2017}
{Axelle Amon et al.}
\newblock Preface: Focus on imaging methods in granular physics.
\newblock {\em Rev. Sci. Instr.}, 88:051701, 2017.

\bibitem{Tsai2003}
J.-C. Tsai, G.~A. Voth, and J.~P. Gollub.
\newblock {Internal Granular Dynamics, Shear-Induced Crystallization, and
  Compaction Steps}.
\newblock {\em Phys. Rev. Lett.}, 91:064301, 2003.

\bibitem{Tsai2004}
J.-C. Tsai and J.~P. Gollub.
\newblock {Slowly sheared dense granular flows: Crystallization and nonunique
  final states}.
\newblock {\em Phys. Rev. E}, 70:031303, 2004.

\bibitem{Panaitescu2012}
Andreea Panaitescu, K.~Anki Reddy, and Arshad Kudrolli.
\newblock Nucleation and crystal growth in sheared granular sphere packings.
\newblock {\em Phys. Rev. Lett.}, 108:108001, Mar 2012.

\bibitem{Dijksman2017}
J.~A. Dijksman, N.~Brodu, and R.~P. Behringer.
\newblock Refractive index matched scanning and detection of soft particles.
\newblock {\em Rev. Sci. Instrum.}, 88:051807, 2017.

\bibitem{Richard2003}
Patrick Richard, Pierre Philippe, Fabrice Barbe, St\'ephane Bourl\`es, Xavier
  Thibault, and Daniel Bideau.
\newblock Analysis by x-ray microtomography of a granular packing undergoing
  compaction.
\newblock {\em Phys. Rev. E}, 68:020301, 2003.

\bibitem{Zhang2006}
W.~Zhang, K.A. Thompson, A.H. Reed, and L.~Beenken.
\newblock Relationship between packing structure and porosity in fixed beds of
  equilateral cylindrical particles.
\newblock {\em {Chem. Eng. Sci.}}, {61}:{8060}, {2006}.

\bibitem{Borzsonyi2012}
T.~B\"orzs\"onyi, B.~Szab\'o, G.~T\"or\"os, S.~Wegner, J.~T\"or\"ok, E.~Somfai,
  T.~Bien, and R.~Stannarius.
\newblock Orientational order and alignment of elongated particles induced by
  shear.
\newblock {\em Phys. Rev. Lett}, 108:228302, 2012.

\bibitem{Wegner2012}
Sandra Wegner, Tamas B\"orzs\"onyi, Tomasz Bien, Georg Rose, and Ralf
  Stannarius.
\newblock Alignment and dynamics of elongated cylinders under shear.
\newblock {\em Soft Matter}, 8:10950--10958, 2012.

\bibitem{Wegner2014}
S.~Wegner, R.~Stannarius, A.~Boese, G.~Rose, B.~Szabo, E.~Somfai, and
  T.~B\"orzs\"onyi.
\newblock Effects of grain shape on packing and dilatancy of sheared granular
  materials.
\newblock {\em Soft Matter}, 10:5157, 2014.

\bibitem{Szabo2014}
B.~Szab\'o, J.~T\"or\"ok, E.~Somfai, S.~Wegner, R.~Stannarius, A.~B\"ose,
  G.~Rose, F.~Angenstein, and T.~B\"orzs\"onyi.
\newblock Evolution of shear zones in granular materials.
\newblock {\em Phys. Rev. E}, 90:032205, 2014.

\bibitem{Stannarius2019}
R.~Stannarius, D.~Sancho Martinez, T.~Finger, E.~Somfai, and T.~B\"orzs\"onyi.
\newblock Packing and flow profiles of soft grains in 3d silos reconstructed
  with x-ray computed tomography.
\newblock {\em Granular Matter}, 21:56, 2019.

\bibitem{Stannarius2019b}
Ralf Stannarius, Diego~Sancho Martinez, Tam\'as B\"orzs\"ony, Martina Bieberle,
  Frank Barthel, and Uwe Hampel.
\newblock High-speed x-ray tomography of silo discharge.
\newblock {\em New J. Phys.}, 21:113054, 2019.

\bibitem{Weis2017}
S.~Weis and M.~Schr\"oter.
\newblock Analyzing x-ray tomographies of granular packings.
\newblock {\em Rev. Sci. Instr.}, 88:051809, 2017.

\bibitem{Sederman2007}
A.~J. Sederman, L.~F. Gladden, and M.~D. Mantle.
\newblock Application of magnetic resonance imaging techniques to particulate
  systems,.
\newblock {\em Adv. Powder Techn.}, 18:23--38, 2007.

\bibitem{Stannarius2017}
R.~Stannarius.
\newblock Magnetic resonance imaging of granular materials.
\newblock {\em Rev. Sci. Instr.}, 88:051806, 2017.

\bibitem{Nakagawa1993}
M.~Nakagawa, S.~A. Altobelli, A.~Caprihan, E.~Fukushima, and E.-K. Jeong.
\newblock Non-invasive measurements of granular flows by resonance imaging.
\newblock {\em Exp. Fluids}, 16:54, 1993.

\bibitem{Nakagawa1994}
M.~Nagakawa.
\newblock Axial segregation of granular flows in a horizontal rotating
  cylinder.
\newblock {\em Chem. Eng. Sci.}, 9:2540, 1994.

\bibitem{Sommier2001}
N.~Sommier, P.~Porion, P.~Evesque, B.~Leclerc, P.~Tchoreloff, and G.~Couarraze.
\newblock Magnetic resonance imaging investigation of the mixing-segregation
  process in a pharmaceutical blender.
\newblock {\em Int. J. Pharmaceutics}, 222:243--258, 2001.

\bibitem{Penn2019}
Alexander Penn, C.M. Boyce, K.P. Pruessmann, and C.R. M\"uller.
\newblock Regimes of jetting and bubbling in a fluidized bed studied using
  real-time magnetic resonance imaging.
\newblock {\em Chemical Engineering Journal}, 383:123185, 10 2019.

\bibitem{Penn2020}
Alexander Penn, A.~Padash, Maxim Lehnert, K.~Pruessmann, C.~M\"uller, and
  Christopher Boyce.
\newblock Asynchronous bubble pinch-off pattern arising in fluidized beds due
  to jet interaction: A magnetic resonance imaging and computational modeling
  study.
\newblock {\em Phys. Rev. Fluids}, 5, 09 2020.

\bibitem{Finger2006}
T.~Finger, A.~Voigt, J.~Stadler, H.G. Niessen, L.~Naji, and Ralf Stannarius.
\newblock Coarsening of axial segregation patterns of slurries in a
  horizontally rotating drum.
\newblock {\em Phys. Rev. E}, 74:031312, 2006.

\bibitem{Naji2009}
L.~Naji and R.~Stannarius.
\newblock {Axial and radial segregation of granular mixtures in a rotating
  spherical container}.
\newblock {\em Phys. Rev. E}, 79:031307, 2009.

\bibitem{Fischer2009}
D.~Fischer, T.~Finger, F.~Angenstein, and R.~Stannarius.
\newblock {Diffusive and subdiffusive axial transport of granular material in
  rotating mixers}.
\newblock {\em Phys. Rev. E}, 80:061302, 2009.

\bibitem{Sakaie2008}
K.~Sakaie, D.~Fenistein, T.~J. Carroll, M.~van Hecke, and P.~Umbanhowar.
\newblock Mr imaging of reynolds dilatancy in the bulk of smooth granular
  flows.
\newblock {\em EPL (Europhysics Letters)}, 84:38001, 2008.

\bibitem{Borzsonyi2011}
T.~B\"orzs\"onyi, T.~Unger, B.~Szab\'o, S.~Wegner, F.~Angenstein, and
  R.~Stannarius.
\newblock Reflection and exclusion of shear zones in inhomogeneous granular
  materials.
\newblock {\em Soft Matter}, 7:8330, 2011.

\bibitem{Engel2020}
Maria Engel, Lars Kasper, Bertram Wilm, Benjamin Dietrich, Laetitia Vionnet,
  Franciszek Hennel, Jonas Reber, and Klaas Pruessmann.
\newblock Mono-planar t-hex: Speed and flexibility for high-resolution 3d
  imaging.
\newblock {\em Mag. Res. Medicine}, 85:2507, 2020.

\bibitem{Bieberle2016}
M.~Bieberle and F.~Barthel.
\newblock Magnetic resonance imaging of granular materials.
\newblock {\em Chem. Eng. Journal}, 285:218, 2016.

\bibitem{Waktola2018}
Selam Waktola, André Bieberle, Frank Barthel, Martina Bieberle, Uwe Hampel,
  Krzysztof Grudzien, and Laurent Babout.
\newblock A new data-processing approach to study particle motion using
  ultrafast x-ray tomography scanner: case study of gravitational mass flow.
\newblock {\em Experiments in Fluids}, 59, 03 2018.

\bibitem{Panaitescu2010b}
Andreea Panaitescu and Arshad Kudrolli.
\newblock Experimental investigation of cyclically sheared granular particles
  with direct particle tracking.
\newblock {\em Progr. Theor. Phys. Suppl.}, 184:100, 2010.

\bibitem{Kuwano2013}
Osamu Kuwano, Ryosuke Ando, and Takahiro Hatano.
\newblock Crossover from negative to positive shear rate dependence in granular
  friction.
\newblock {\em Geophys. Res. Lett.}, 40:1295--1299, 2013.

\bibitem{Fall2015}
Abdoulaye Fall, Guillaume Ovarlez, David Hautemayou, C\'edric M\'ezi\`{e}re,
  Jean-No\"el Roux, J.-N. Roux, and F.~Chevoir.
\newblock Dry granular flows: Rheological measurements of the $\mu$(i)
  -rheology.
\newblock {\em J. Rheology}, 59:1065, 2015.

\bibitem{Fenistein2003}
Denis Fenistein and Martin van Hecke.
\newblock Wide shear zones in granular bulk flow.
\newblock {\em Nature}, 425:256, 2003.

\bibitem{Fenistein2004}
Denis Fenistein, Jan-Willem van~de Meent, and Martin van Hecke.
\newblock Universal and wide shear zones in granular bulk flow.
\newblock {\em Phys. Rev. Lett.}, 92:094301, 2004.

\bibitem{Cheng2006}
Xiang Cheng, Jeremy~B. Lechman, Antonio Fernandez-Barbero, Gary~S. Grest,
  Heinrich~M. Jaeger, Greg~S. Karczmar, Matthias~E. M\"obius, and Sidney~R.
  Nagel.
\newblock Three-dimensional shear in granular flow.
\newblock {\em Phys. Rev. Lett.}, 96:038001, 2006.

\bibitem{Fenistein2006}
Denis Fenistein, Jan-Willem van~de Meent, and Martin van Hecke.
\newblock Core precession and global modes in granular bulk flow.
\newblock {\em Phys. Rev. Lett.}, 96:118001, Mar 2006.

\bibitem{Unger2004}
T.~Unger, J.~T\"or\"ok, J.~Kert\'esz, and D.~E. Wolf.
\newblock Shear band formation in granular media as a variational problem.
\newblock {\em Phys. Rev. Lett.}, 92:214301, May 2004.

\bibitem{Unger2007}
J.~T\"or\"ok, T.~Unger, J.~Kert\'esz, and D.~E. Wolf.
\newblock Shear zones in granular materials: Optimization in a self-organized
  random potential.
\newblock {\em Phys. Rev. E}, 75:011305, Jan 2007.

\bibitem{Dijksman2010}
Joshua~A. Dijksman and Martin van Hecke.
\newblock Granular flows in split-bottom geometries.
\newblock {\em Soft Matter}, 6:2901, 2010.

\bibitem{Wortel2015}
G.~Wortel, T.~B\"orzs\"onyi, E.~Somfai, S.Wegner, B.~Szab\'o, R.~Stannarius,
  and M.~van Hecke.
\newblock Heaping, secondary flows and broken symmetry in flows of elongated
  granular particles.
\newblock {\em Soft Matter}, 11:2570, 2015.

\bibitem{Fischer2016}
D.~Fischer, T.~B\"orzs\"onyi, D.~Nasato, T.~P\"oschel, and R.~Stannarius.
\newblock Heaping and secondary flows in sheared granular materials.
\newblock {\em New J. Phys.}, 18:113006, 2016.

\bibitem{Ashour2017b}
A.~Ashour, T.~Trittel, T.~B\"orzs\"onyi, and R.~Stannarius.
\newblock Silo outflow of soft frictionless spheres.
\newblock {\em Phys. Rev. Fluids}, 2:123302, 2017.

\bibitem{Kjaer1987}
L.~Kjaer, C.~Thomsen, O.~Henriksen, P.~Ring, M.~Stubgaard, and E.~J. Pedersen.
\newblock Evaluation of relaxation time measurements by magnetic resonance
  imaging.
\newblock {\em Acta Radiologica}, 28:345, 1987.

\bibitem{Harth2020}
Kirsten Harth, Jing Wang, Tamas B\"orzs\"onyi, and Ralf Stannarius.
\newblock Intermittent flow and transient congestions of soft spheres passing
  narrow orifices.
\newblock {\em Soft Matter}, 16:8013, 2020.

\bibitem{Wang2021a}
Jing Wang, Kirsten Harth, Ralf Stannarius, Bo~Fan, and Tam\'as B\"orzs\"onyi.
\newblock Discharge of soft and hard grains and their mixtures from {2D} silos.
\newblock {\em EPJ Web of Conferences}, 249:03002, 2021.

\bibitem{Wang2021}
Jing Wang, Bo~Fan, Tivadar Pong\'o, Kirsten Harth, Torsten Trittel, Ralf
  Stannarius, Maja Illig, Tam\'as B\"orzs\"onyi, and Ra\'ul~Cruz Hidalgo.
\newblock Silo discharge of mixtures of soft and rigid grains.
\newblock {\em Soft Matter}, 17:4282, 2021.

\bibitem{Pongo2021}
Tivadar Pong\'o, J\'anos T\"or\"ok, Vikt\'oria Stiga, Bal\'azs Szab\'o, S\'ara
  L\'evay, Ralf Stannarius, Ra\'ul~Cruz Hidalgo, and Tam\'as B\"orzs\"onyi.
\newblock Flow in an hourglass: particle friction and stiffness matter.
\newblock {\em New J. Phys.}, 23:023001, 2021.

\end{thebibliography}
\end{document}